\documentclass[aps,prb,twocolumn,groupedaddress,showpacs]{revtex4}

\usepackage{psfrag}

\usepackage{graphicx}
\usepackage{amsmath}
\usepackage{amssymb}
\usepackage{bm}     

\newcommand{\ban}{\begin{eqnarray*}}
\newcommand{\ean}{\end{eqnarray*}}
\newcommand{\bea}{\begin{eqnarray}}
\newcommand{\eea}{\end{eqnarray}}

\newcommand{\rd}[1]{\mathop{\mathrm{d}#1}} 

\newcommand{\norm}[1]{\left\vert #1\right\vert}

\newcommand{\av}[1]{\left\langle#1\right\rangle}

\newcommand\Tr{\text{tr}}
\newcommand\Trw{\Tr\left[(\win)^2\right]}

\newcommand\figwidth{3.375in}

\providecommand{\openone}{\leavevmode\hbox{\small1\kern-3.8pt\normalsize1}}

\newcommand\NL{\ensuremath{N}}
\newcommand\NR{\ensuremath{M}}
\newcommand\Ntot{\ensuremath{K}}
\newcommand\Ntotp{\ensuremath{\Ntot_\phi}}

\newcommand\dendef{\Ntot (2\Ntot-1) (2\Ntot-3)}

\newcommand\den{\Lambda}
\newcommand\denp{\den_\phi}
\newcommand\Nphi{\ensuremath{N_\phi}}
\newcommand\win[1][]{\ensuremath{w^{\text{in}#1}}}
\newcommand\wout{\ensuremath{w^{\text{out}}}}

\newcommand\cur{\ensuremath{g^s}}
\providecommand\prob{\ensuremath{g}} 
\providecommand\pol{\ensuremath{p}} 

\providecommand\epp{\epsilon^\prime}
\providecommand\fermi{f}

\graphicspath{{Figs/}}

\begin{document}
\title{Spin polarized current generation from quantum dots
without magnetic fields}
\author{Jacob J.\ Krich}
\author{Bertrand I.\ Halperin}
\affiliation{Physics Department, Harvard University, Cambridge,
MA 02138}
\date{\today}

\begin{abstract}
  An unpolarized charge current passing through a chaotic
  quantum dot with spin-orbit coupling can produce a spin
  polarized exit current without magnetic fields or
  ferromagnets. We use random matrix theory to estimate the
  typical spin polarization as a function of the number of
  channels in each lead in the limit of large spin-orbit
  coupling. We find rms spin polarizations up to 45\% with one
  input channel and two output channels. Finite temperature and
  dephasing both suppress the effect, and we include dephasing
  effects using a new variation of the third lead model. If
  there is only one channel in the output lead, no spin
  polarization can be produced, but we show that dephasing
  lifts this restriction.
\end{abstract}
\pacs{72.25.-b,73.63.Kv,75.47.-m,85.75.-d}
 \maketitle

\section{Introduction}

The generation and control of spin polarized currents, in
particular without magnetic fields or ferromagnets, is a major
focus of recent experimental and theoretical work. This
includes the spin Hall effect, which produces spin currents
transverse to an electric field in a two-dimensional electron
system (2DES) with spin-orbit coupling, with spin accumulation
at the edges. \cite{Engel06} Similarly, the magnetoelectric
effect \cite{Levitov85,edelstein1990,aronov91} produces a
steady state spin accumulation when an electric field is
applied to a 2DES with spin-orbit coupling. The accumulation
can be uniform \cite{trushin07,huang06} in the case of uniform
Rashba spin-orbit coupling \cite{Bychkov84} or at the edges of
a channel in either the Rashba model
\cite{Chesi07,Chaplik02,Liu07} or with spin-orbit coupling
induced by lateral confinement. \cite{jiang06,Xing06}
Experiments have observed current induced spin polarization in
$n$-type 3D samples \cite{Kato04} and in 2D hole systems
\cite{Ganichev06,silov04} with spin polarization estimated to
be up to 10\%. \cite{silov04} Further work suggests a spin
polarized current can be produced by a quantum point contact
(QPC) with spin-orbit coupling, \cite{Silvestrov06,Eto05} in a
carbon nanotube, \cite{Jiang07} in a ballistic ratchet,
\cite{Scheid07} in a torsional oscillator, \cite{Kovalev07} or
in vertical transport through a quantum well.
\cite{Malshukov07}

Here we show that generating a polarized current from an
unpolarized current is a generic property of scattering through
a mesoscopic system with spin-orbit coupling. We propose using
many-electron quantum dots (outside the Coulomb blockade
regime) with spin-orbit coupling to produce partially spin
polarized currents without magnetic fields or ferromagnets. Due
to the complicated boundary conditions of the quantum dot, we
do not solve for the spin polarization in terms of any
particular spin-orbit coupling model, geometry, and contact
configuration. We estimate the effect for a ballistic system in
the limit of strong spin-orbit coupling by performing a random
matrix theory (RMT) calculation for the spin polarization,
allowing consideration of realistic quantum dot devices robust
to details of shape and contact placement. Finely tuned systems
should be able to exceed these polarizations, but these results
provide a useful benchmark for whether a particular tuned
system is better than a generic chaotic one. We use a density
matrix formalism throughout, which allows us to develop
straightforwardly a spin-conserving dephasing probe, using a
new variant of the third-lead technique for accounting for
dephasing. Dephasing and finite temperature both reduce the
expected polarization. Without dephasing, we find that if there
is only one outgoing channel then no spin polarization is
possible, which was first shown by Zhai and Xu. \cite{Zhai05}
Interestingly, with dephasing, spin polarization can be
produced with only one outgoing channel. The case of polarized
input currents will be discussed elsewhere.
\cite{krichUnpublished}

Analogous calculations have been performed by Bardarson,
Adagideli, and Jacquod in a four-terminal geometry, to study
the transverse spin current produced by an applied charge
current. \cite{Bardarson07}

\section{Setup and Symmetry Restrictions}
 \begin{figure}
    \psfrag{up}{$\uparrow$}%
    \psfrag{do}{$\downarrow$}%
   \includegraphics[width=2.25in]{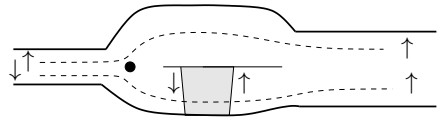}
  \caption{A tuned model quantum dot with $\NL=1$ channels in
  the left lead and $\NR=2$ channels in the right lead. A skew
  scatterer sends all $\hat z$ spins to the top channel and all
  $-\hat z$ spins to the bottom channel. The shaded area in the
  bottom channel has Rashba spin-orbit coupling of precisely
  the strength to rotate a down spin at the Fermi energy to an
  up spin, thus producing a spin polarized exit current from
  any input current, while respecting time reversal symmetry.
  \label{fig:QD}}
\end{figure}

We consider non-interacting electrons in a quantum dot with two
attached leads connected to large reservoirs. For any electron
current entering from the leads, we can describe the output
state in the leads in terms of the S-matrix of the dot,
including any tunnel barriers between the leads and the dot. We
assume negligible spin-orbit coupling in the leads and consider
the lead on the left (right) to have $\NL$ ($\NR$)
spin-degenerate channels at least partially open at the Fermi
energy, and let $\Ntot=\NL+\NR$. As usual, the channel
wavefunctions are normalized so all channels have the same
flux. The S-matrix $S$ is a $2\Ntot \times 2\Ntot$ unitary
matrix of complex numbers. For spin 1/2 particles with
spin-orbit coupling, however, it is convenient to consider $S$
to be a $\Ntot\times \Ntot$ matrix of quaternions. We give a
brief introduction to quaternions and explain why they are
easier to work with.

We choose a representation of the quaternions such that a
quaternion $q$ is a $2\times2$ matrix of complex numbers
   $ q=q^{(0)}\openone_2+i\sum_{\mu=1}^3 q^{(\mu)}\sigma_\mu$,
where $\sigma_\mu$ are the Pauli matrices and $q^{(\mu)} \in
\mathbb{C}$.
We define three conjugates of $q$: complex conjugate
$q^*=q^{(0)*}\openone_2+i\sum q^{(\mu)*}\sigma_\mu$, quaternion
dual $q^R=q^{(0)}\openone_2-i\sum q^{(\mu)}\sigma_\mu$, and
Hermitian conjugate $q^\dagger=q^{R*}$. For a
$\Ntot\times\Ntot$ matrix of quaternions, $Q$, we define
  \mbox{$\left(Q^*\right)_{i j}=\phantom{(}Q_{i j}^{\phantom{i j} *}$},
  $\left(Q^R\right)_{i j}=\phantom{(}Q_{j i}^{\phantom{j i} R}$,
  and $\left(Q^\dagger\right)=\left(Q^*\right)^R$.
By convention, the trace of the quaternion matrix $Q$ is
  $\Tr Q= \sum_i Q_{i i}^{(0)}$.

Given a $\Ntot\times\Ntot$ quaternion matrix $Q$, we can
associate it with a $2\Ntot\times2\Ntot$ complex matrix $A$ in
the obvious way.  Note then that $Q^\dagger$ is equivalent to
$A^\dagger$, the usual Hermitian conjugate of a complex matrix,
but $Q^*$ is not equivalent to $A^*$.  Note further that the
trace convention implies that $\Tr A=2\Tr Q$.

The quaternion representation is convenient, as the time
reversal operation for a scattering matrix can simply be
written as $S\rightarrow S^R$. \cite{Mehta04} The S-matrix of a
system with time reversal symmetry (TRS) is \emph{self-dual}.

If \win{} (\wout) is the $\Ntot\times\Ntot$ quaternion density
matrix of the incoming (outgoing) current, $\wout=S\win
S^\dagger$. The density matrix describing the unpolarized
incoherent combination of all \NL{} incoming channels is
\begin{eqnarray}
    \win= \frac{1}{2\NL}\begin{pmatrix}
                         \openone_\NL & \\
                                  & 0_\NR
                         \end{pmatrix}
  \label{eq:win_unpolarized}
\end{eqnarray}
That is, $\win=P_L/2\NL$ where $P_L$ is the projection onto the
channels of the left lead. We choose $\Tr \win=1/2$, due to the
trace convention.

The Landauer-B\"uttiker formula gives the conductance in terms
of the S-matrix. \cite{Buttiker85} We write the
$\Ntot\times\Ntot$ quaternion S-matrix as %
$\left(\begin{smallmatrix}
    r &t^\prime\\
    t &r^\prime
\end{smallmatrix}\right)$
with $r$ ($r^\prime$) being the $\NL\times\NL$ ($\NR\times\NR$)
reflection matrix and $t$ ($t^\prime$) the $\NR\times\NL$
($\NL\times\NR$) transmission matrix. Then we write the
Landauer-B\"uttiker formula in units of $2e^2/h$ as
\begin{align}
  G &=\Tr(t t^\dagger),\\\nonumber
    &=\Tr(P_R S P_L S^\dagger),\\\nonumber
    &=2\NL\Tr(P_R S \win S^\dagger),\\\nonumber
    &=2\NL\Tr(P_R \wout)\nonumber,
\end{align}
where $P_R$ is the projection onto the channels of the right
lead. Since \win{} is normalized to represent one input
particle entering the system, $\prob=2\Tr(P_R \wout)$ is the
probability for that particle to exit through the right lead.
The conductance is $\NL$ times this probability, so we call
$\prob$ the conductance per channel in the left lead.

Similarly, we define a vector spin conductance \cite{Zhai05}
(i.e., exit spin current divided by voltage) $\vec G^s$ in
units of $e/2\pi$ as
\begin{align}
  \vec G^s&=\Tr(\vec\sigma t t^\dagger),\\\nonumber
          &=2\NL\Tr(\vec\sigma P_R \wout).
\end{align}
Then
\begin{align}\label{eq:def_n}
  \vec\cur=2\Tr(\vec\sigma P_R \wout)
\end{align}
is the spin conductance per channel in the left lead. Hence,
$\cur_\mu$ is the $\mu$-component of the spin polarization of
the exit current times the probability of exiting into the
right lead. Thus, the spin polarization of the current in the
right lead is $\vec\pol=\vec\cur/\prob$, with
$\norm{\pol}\le1$.

We can, of course, construct $\prob$, $\vec\cur$, and
$\vec\pol$ using only the S-matrix and not the density matrices
\win{} and \wout{}. The density matrix approach, however, gives
the flexibility to consider arbitrarily correlated states of
incoming current and also to look for arbitrary correlations in
the outgoing current. \cite{krichUnpublished} We will also use
it to straightforwardly derive a method of accounting for
non-magnetic dephasing in a device with spin-orbit coupling. To
complete the translation to the standard notation of
conductances, we consider sending up- or down-polarized
electrons into a sample and collecting either up- or
down-polarized electrons, giving a conductance matrix
\cite{nikolic05}
\begin{align}
  \mathbf{G}=\begin{pmatrix}
    G_{\uparrow\uparrow}&G_{\uparrow\downarrow}\\
    G_{\downarrow\uparrow}&G_{\downarrow\downarrow}
  \end{pmatrix}
\end{align}
with the total charge conductance being
$G=G_{\uparrow\uparrow}+G_{\uparrow\downarrow}+
G_{\downarrow\uparrow} +G_{\downarrow\downarrow}$.
$G_{\sigma,\sigma^\prime}$ is the conductance for an input
current of spin $\sigma^\prime$ and an exit current of spin
$\sigma$, for $\sigma,\sigma^\prime=\uparrow,\downarrow$.  We
translate the quaternion representation into the standard
notation by noting that the up-polarized incoherent input
current has input density matrix
$\win_\uparrow=\frac{1+\sigma_3}{2\NL}P_L$. The output density
matrix is $\wout_\uparrow=S\win_\uparrow S^\dagger$ and the
portion representing the output in the right lead is
$t\frac{1+\sigma_3}{2\NL}t^\dagger$. The Landauer-B\"uttiker
formula gives, in units of $2e^2/h$,
\begin{align}
    G_{\uparrow\uparrow}
    =&\NL\Tr(P_R\tfrac{1+\sigma_3}{2}\wout_\uparrow),\nonumber\\
    =&\Tr(\tfrac{1+\sigma_3}{2}t\tfrac{1+\sigma_3}{2}t^\dagger).
\intertext{Similarly,}
  G_{\downarrow\uparrow}=&\Tr(\tfrac{1-\sigma_3}{2}t\tfrac{1+\sigma_3}{2}t^\dagger),\\
  G_{\uparrow\downarrow}=&\Tr(\tfrac{1+\sigma_3}{2}t\tfrac{1-\sigma_3}{2}t^\dagger),\\
  G_{\downarrow\downarrow}=&\Tr(\tfrac{1-\sigma_3}{2}t\tfrac{1-\sigma_3}{2}t^\dagger),
\end{align}
from which we see that $G=\Tr(t t^\dagger)$, which is the usual
Landauer-B\"uttiker formula. \cite{Buttiker85}

Though there are several spin-orbit coupled systems that
demonstrate spin polarization from unpolarized input, in many
cases the effect is subtle.
\cite{Silvestrov06,Eto05,Jiang07a,Scheid07} Here we give an
idealized thought experiment showing that an unpolarized input
current can produce a spin polarized output current.
Consider a system with $\NL=1$ and $\NR=2$, as illustrated in
Fig.\ \ref{fig:QD}.  All input electrons are incident on a
perfect skew scatterer which sends spins quantized in the $+z$
direction into exit channel 1 and spins quantized in the $-z$
direction into exit channel 2. Exit channel 2 has a region with
Rashba spin-orbit coupling \cite{Bychkov84} which is precisely
of the strength and length necessary to rotate $-z$ spins to
$+z$. Thus, all spins incident from the left lead exit with
their spins up, and the system respects TRS, since skew
scattering and Rashba spin-orbit interaction are each time
reversal symmetric.

We illustrate by constructing $S$ explicitly. We can express
the scattering matrix for this thought experiment (up to an
overall phase) in the $6\times6$ and $3\times3$ representations
as
\begin{alignat}{10}
  S&=&&\left(\begin{smallmatrix}
    0 & 0 & 0 & 0 & 0 & -1\\
    0 & 0 & 0 & 1 & 0 & 0 \\
    1 & 0 & 0 & 0 & 0 & 0 \\
    0 & 0 & 0 & 0 & e^{i \theta} & 0 \\
    0 & 1 & 0 & 0 & 0 & 0 \\
    0 & 0 & -e^{i\theta}& 0 & 0 & 0
  \end{smallmatrix}\right)\\
  &\equiv \frac{1}{2}&&\left(\begin{smallmatrix}
    0           & 1-\sigma_z    & -\sigma_x-i\sigma_y\\
    1+\sigma_z  & 0             & e^{i\theta} (\sigma_x-i\sigma_y) \\
    \sigma_x+i\sigma_y & -e^{i\theta}(\sigma_x-i\sigma_y) &0\\
  \end{smallmatrix}\right)
\end{alignat}
where $\theta\in[0,2\pi)$ and $r$ and $t$ have been determined
by the above description, while the rest of the matrix is given
by TRS and unitarity. The unpolarized input quaternion density
matrix is
$\win=\left(\begin{smallmatrix}1/2&&\\&0&\\&&0\end{smallmatrix}\right)$,
giving
\begin{align}
    \wout=S\win S^\dagger=\frac{1}{4}
    \begin{pmatrix}
      0 & 0          & 0\\
      0 & 1+\sigma_z & 0 \\
      0 & 0          & 1+\sigma_z
    \end{pmatrix},
\end{align}
so Eq.\ \ref{eq:def_n} gives $\vec{\cur}=\vec\pol=\hat{z}$, as
stated above.

We now prove that having at least two channels in the outgoing
lead is essential. That is, for a dot with TRS and $\Ntot$
channels in attached leads, if an unpolarized equally weighted
incoherent current is sent into $\NL=\Ntot-1$ of the channels,
then the spin polarization in the remaining channel must be
zero. This result has been shown before, \cite{Zhai05} but the
quaternion formalism with density matrices makes it
particularly transparent, so we include the proof here.

We start with
\begin{align}
  \win=\frac{1}{2\NL}\begin{pmatrix}
    \openone_\NL\\
    &0
  \end{pmatrix}
  =\frac{\openone_\Ntot-P_\Ntot}{2\NL},
\end{align}
where $P_\Ntot$ is the projection onto the $\Ntot^\text{th}$
channel. The quaternion scattering matrix satisfies $S=S^R$
since TRS is unbroken, and
\begin{align}
    \wout=\frac{S S^\dagger - S P_\Ntot
            S^\dagger}{2\NL}=\frac{\openone_\Ntot-S P_\Ntot S^\dagger}{2\NL}.
\end{align}
Note that $S=S^R$ implies both $S^\dagger=S^*$ and $S_{i i} \in
\mathbb{C}$ for $i=1\ldots\Ntot$.

Using Eq.\ \ref{eq:def_n}, the spin conductance is
$\cur_\mu=2i[\wout_{\Ntot \Ntot}]^{(\mu)}$. In particular, if
$\wout_{\Ntot\Ntot}$ has no quaternion part, then $\cur_\mu=0$.
We have
\begin{align}
  \wout_{\Ntot\Ntot}=\frac{1-S_{\Ntot\Ntot}S^*_{\Ntot\Ntot}}{2\NL},
\end{align}
and $S_{\Ntot\Ntot} S^*_{\Ntot\Ntot}$ is real, so
$\wout_{\Ntot\Ntot}\in\mathbb{R}$ and $\vec\cur=0$. This proof
applies with channels that are fully open or have tunnel
barriers, as it requires only that the S-matrix satisfy TRS and
unitarity, which are unchanged by tunnel barriers.

We note further that if $\Ntot>2$ then 1) the \emph{reflected}
current in any of the $\Ntot-1$ input channels can be spin
polarized, and 2) if the input current goes through less than
$\Ntot-1$ channels, then the remaining channels can have a spin
polarization, as shown in the example of Fig.\ \ref{fig:QD}.

\section{Random Matrix Theory}

We estimate the expected spin polarization in realistic
situations by using random scattering matrix theory. We assume
that the mean dwell time $\tau_d$ of particles in the dot is
much greater than the Heisenberg time
$\tau_H=2\pi\hbar/\Delta$, where $\Delta=2\pi\hbar^2/m A$ is
the mean orbital level spacing, $m$ is the effective mass, and
$A$ is the area of the dot. We further assume the strong
spin-orbit limit, where the spin-orbit time $\tau_\text{so}$ is
much less than $\tau_d$. For a chaotic quantum dot, $\tau_d=m
A/\hbar\Ntot$, where $\Ntot$ is the number of fully open
orbital channels attached to the dot, so for a sufficiently
large $A$, even a material with ``weak'' spin-orbit coupling
will be in the strong spin-orbit limit. The crossover from weak
to strong spin-orbit coupling in chaotic quantum dots has been
studied in the $\Ntot\gg1$ limit in the context of adiabatic
spin pumping. \cite{Sharma03}

For dots with strong spin-orbit coupling, we assume that the
S-matrix is chosen from the uniform distribution of unitary
matrices subject to TRS, called the circular symplectic
ensemble (CSE). \cite{Mehta04,beenakker97} We find the root
mean square (rms) magnitude of the spin conductance on
averaging over the CSE, which gives the typical spin
conductance magnitude to be expected from chaotic devices. Such
an averaging can be realized in practice by small alterations
of the dot shape. \cite{zumbuhl02,zumbuhl05}

By symmetry, $\av{\cur_\mu}=0$ for $\mu=1,2,3$. Using Eq.\
\ref{eq:def_n}, we evaluate
\begin{align}
  \av{(\cur)^2}=4\av{\Tr(\sigma_\mu P_R S\win S^\dagger)
                \Tr(\sigma_\mu P_R S\win S^\dagger)},
\end{align}
where we sum over $\mu$.

We use the technique for averaging over the CSE described by
Brouwer and Beenakker in Section V of
Ref.~\onlinecite{brouwer96}. We need just two generic averages,
which we will use repeatedly. The first is of the form
$\av{F_1(S)}=\av{\Tr(A S B S^\dagger)}$, where $A$ and $B$ are
constant $\Ntot\times\Ntot$ quaternion matrices and the average
is taken over $S$ chosen from the CSE of $\Ntot\times\Ntot$
quaternion self-dual matrices.  Then\cite{brouwer96}
\begin{align} \label{eq:2S}
  \av{F_1}=\frac{1}{2\Ntot -1}[2\Tr(A)\Tr(B)-\Tr(A B^R)].
\end{align}
The second average we need is $\av{F_2(S)}=\av{\Tr(A S B
S^\dagger)\Tr(C S D S^\dagger)}$ where $A$, $B$, $C$, $D$ are
constant $\Ntot\times\Ntot$ quaternion matrices and $A B=A D=C
B=C D=0$. We find\cite{brouwer96}
\begin{alignat}{10}\label{eq:4S}
  \av{F_2}&=\frac{1}{\den}&\big\lbrace
        &(\Ntot-1)[4\Tr A\Tr B\Tr C\Tr D+\Tr(AC)\Tr(BD)]\nonumber\\
        &&&-[\Tr A\Tr C\Tr(BD)+\Tr(AC)\Tr B\Tr D]\big\rbrace,
\end{alignat}
where $\den=\dendef$.

Using Eq.\ \ref{eq:4S}, we find
\begin{eqnarray}
  \av{(\cur)^2}&=& 3\frac{\NR(\NR-1)}{\NL\den},
            \label{eq:n_msq_unpolarized}
\end{eqnarray}
where we used $\Tr(\sigma_\mu P_R)=0$, $\Tr(P_R^2)=\NR$,
$\Tr\win=1/2$, and $\Trw=1/4\NL$. Note that when $\NR=1$,
$\av{(\cur)^2}=0$, consistent with the general symmetry.

If we are interested in the mean square polarization of the
exit current, $\av{\pol^2}=\av{(\cur)^2 \prob^{-2}}$, we can
approximate it by $\av{(\cur)^2}/\av{\prob}^2$. This
approximate form is useful for analytical progress and will be
compared to numerical results. Using Eq.\ \ref{eq:2S},
\begin{align}
    \av{\prob}=\frac{2\NR}{2\Ntot-1}, \label{eq:p2}
\end{align}
which, combined with Equation \ref{eq:n_msq_unpolarized}, gives
\begin{eqnarray}
    \av{\pol^2}&\approx&\frac{3(\NR-1)(2\Ntot-1)}
      {4\NR\NL\Ntot(2\Ntot-3)}\label{eq:s2}.
\end{eqnarray}

We study the approximation
$\av{(\cur)^2\prob^{-2}}\approx\av{(\cur)^2}/\av{\prob}^2$
numerically. We choose a $2\Ntot \times 2\Ntot$ complex
Hermitian matrix from the Gaussian unitary ensemble
\cite{Mehta04} and find the unitary matrix $U$ which
diagonalizes it.  We multiply columns of $U$ by random phases,
map $U$ into a $\Ntot\times\Ntot$ matrix of quaternions, and
construct unitary self-dual $S$ by setting $S=U U^R$, giving
$S$ chosen from the CSE. \cite{Forrester05ch2}
\begin{figure}
   \includegraphics[width=\figwidth]{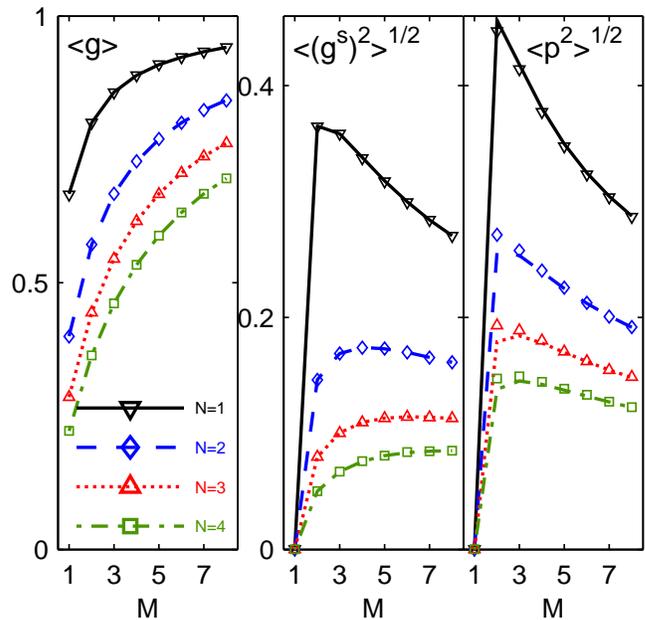}
  \caption{(color online) Numerical (symbols) and analytical
  (lines) results for normalized mean conductance $\av{\prob}$,
  rms spin conductance $\cur$, and rms spin polarization $\pol$
  of current exiting a chaotic quantum dot with $\NL$ ($\NR$)
  channels in the entrance (exit) lead. An average over 60000
  S-matrices from the CSE was performed for each data point.
  The lines are from Eqs.\
  \ref{eq:n_msq_unpolarized}--\ref{eq:s2}. \label{fig:s2}}
\end{figure}
\begin{figure}
   \includegraphics[width=\figwidth]{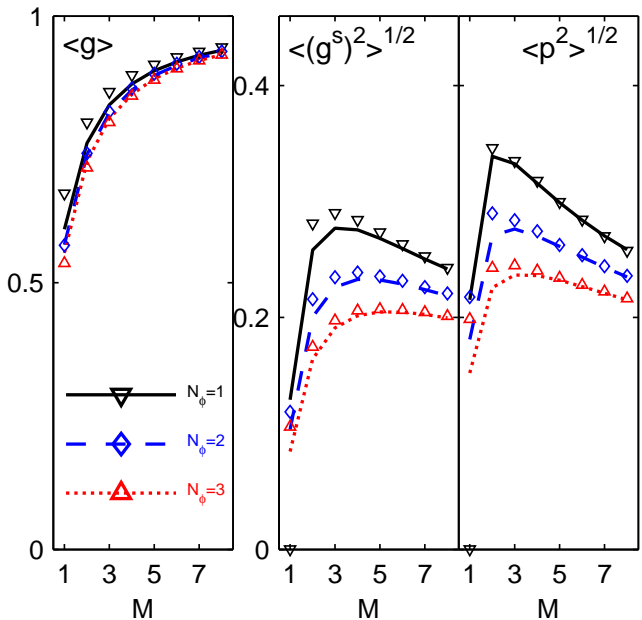}
  \caption{(color online) Numerical (symbols) and analytical
  (lines) results for normalized mean conductance $\av{\prob}$,
  rms spin conductance $\cur$, and rms spin polarization $\pol$
  of current exiting a chaotic quantum dot with $\NL=1$ channel
  in the entrance lead. $\NR$ and $\Nphi$ are the numbers of
  channels in the exit and dephasing leads, respectively. An
  average over 60000 S-matrices from the CSE was performed for
  each data point. The lines are from Eqs.\ \ref{eq:n_phi} and
  \ref{eq:p2_dephasing}.
  \label{fig:s2_phi}}
\end{figure}

 Figure \ref{fig:s2} shows the numerical and analytical
results, which agree quantitatively for $\av{\prob^2}$ and
$\av{(\cur)^2}$ and qualitatively for $\av{\pol^2}$. The
largest percentage disagreement for $\av{\pol^2}$ is 7\%.

\section{Dephasing} We add dephasing to this setup using the
dephasing voltage probe technique.
\cite{buttiker88,brouwer95,baranger95}  We add a fictitious
voltage probe drawing no current with
$\Nphi=2\pi\hbar/\Delta\tau_\phi$ fully open orbital channels,
where $\tau_\phi$ is the dephasing time. We extend this model
to preserve the spin of the reinjected electrons.

In contrast with previous work, we explicitly model reinjection
of electrons from the voltage lead by modifying \win{} to
include incoherent reinjection from the dephasing lead. The
reinjection matches the total charge/spin current absorbed by
the dephasing lead, but distributes the charge/spin current
evenly between the channels, and thus models dephasing
processes that preserve electron spin. First, consider
  $\eta_\mu=\Tr(\sigma_\mu P_\phi S \win_0 S^\dagger)$,
where $\mu=0,1,2,3$, $\sigma_0=\openone_2$, $P_\phi$ is the
projection operator onto the dephasing lead's channels, and
$\win_0$ is the input density matrix. Then $2\eta_0$ is the
probability for a particle to enter the dephasing lead, and
$2\vec\eta$ is the spin conductance into the dephasing lead,
which is proportional to the spin current into the dephasing
lead.

We reinject from the dephasing lead with
\begin{equation}
  w^\phi_1=\begin{pmatrix}
                  0_{\Ntot}             \\
                    &   & c^1_\mu \sigma_\mu \openone_{\Nphi}\\
                \end{pmatrix}
                =c^1_\mu\sigma_\mu P_\phi,
  \label{eq:w1_phi}
\end{equation}
where we sum over repeated index $\mu$. We set
$c^1_\mu=\eta_\mu/\Nphi$, which ensures the reinjected
charge/spin current equals the absorbed charge/spin current.
Some of this reinjected current reflects back into the
dephasing lead, so it must be reinjected again. We define a
$4\times4$ complex matrix $\Theta_{\mu \nu}=\Tr(\sigma_\nu
P_\phi S \sigma_\mu P_\phi S^\dagger)$, which gives the
charge/spin current in the dephasing lead due to this
reinjection. Defining $\win=\win_0+\win_\phi$, this procedure
gives
\begin{eqnarray}
  \win_\phi  &=&\sum_{n=1}^\infty w^\phi_n \nonumber\\
        &=&P_\phi \sigma_\mu
        \Tr(\sigma_\nu P_\phi S \win_0 S^\dagger)
        \sum_{n=1}^\infty
        \frac{(\Theta^{n-1})_{\mu \nu}}{\Nphi^n} \nonumber\\
        &=&P_\phi \sigma_\mu
        \Tr(\sigma_\nu P_\phi S \win_0 S^\dagger)
        (\Nphi \delta_{\mu \nu} - \Theta_{\mu \nu})^{-1},
        \label{eq:win_phi}
\end{eqnarray}
where we sum over repeated indices $\mu, \nu=0,1,2,3$. This
result holds for any input current, not just the unpolarized
incoherent $\win_0$ discussed here.

We approximate $\win_\phi$ by replacing $\Theta_{\mu\nu}$ with
its average in Eq.\ \ref{eq:win_phi}, similar to Eq.\
\ref{eq:s2}. Using Eq.\ \ref{eq:2S},
\begin{equation}
  \av{\Theta_{\mu \nu}}=\frac{\Nphi}{2\Ntotp-1}
  [2(\Nphi-1)\delta_{\mu 0}\delta_{\nu 0}+\delta_{\mu \nu}],\label{eq:Theta}
\end{equation}
where $\Ntotp=\NL+\NR+\Nphi$. We further replace
$\Tr(\sigma_\mu P_\phi S \win S^\dagger)$ by its average,
\begin{equation}
  \av{\Tr(\sigma_\mu P_\phi S \win S^\dagger)}
  =\frac{\delta_{\mu 0}\Nphi}{2\Ntotp-1},\label{eq:Theta2}
\end{equation}
which gives
\begin{equation}
  \win\approx \win_0 +P_\phi/2\Ntot. \label{eq:win_phi_avg}
\end{equation}
 This turns out to be the same result as if we had chosen
$c^1_\mu=0$ in Eq.\ \ref{eq:w1_phi} for $\mu=1,2,3$.  That is,
in the approximation of Eqs.\ \ref{eq:Theta}--\ref{eq:Theta2},
if we have total spin decay in the dephasing lead, then Eq.\
\ref{eq:win_phi_avg} is unchanged. Note that Eq.\
\ref{eq:win_phi_avg} satisfies unitarity only on average; the
total probability of exiting either through the right or left
lead equals 1 only on average.

In this approximation, we use Eq. \ref{eq:4S} to find
\begin{eqnarray}
  \av{(\cur)^2}\approx\frac{3\NR}{\NL \denp} \left( \NR-1 +
    \Nphi\frac{\NR^2+\NL (\NR-1)}{\Ntot^2}\right) \label{eq:n_phi},
\end{eqnarray}
where $\denp=\Ntotp(2\Ntotp-1)(2\Ntotp-3)$.

Note that even if there is only one outgoing channel, $\NR=1$,
the spin conductance is predicted to be nonzero due to
dephasing. The dephasing induced spin conductance is present
for $\Nphi>1$, as shown in numerical simulations in
Fig.~\ref{fig:s2_phi} and is not an artifact of
Eq.~\ref{eq:win_phi_avg}. In the case of $\Nphi=\NR=1$, an
exact treatment shows that $\vec\cur=0$, contrary to
Eq.~\ref{eq:n_phi}, even with arbitrary tunnel barriers between
the leads and the sample. As shown in Fig.~\ref{fig:s2_phi},
Eq.~\ref{eq:n_phi} works well for $\NR>1$ or $\Nphi>1$.

We can modify this model to have $\Nphi$ dephasing leads each
with one channel, each separately reinjecting the same
charge/spin that it absorbs.  In this model, too, a nonzero
$\vec{\cur}$ can be produced for $\Nphi>1$ (not shown).

Brouwer and Beenakker modified the third-lead dephasing model
to make dephasing uniform in phase space by placing a tunnel
barrier with transparency $\Gamma$ between the dephasing lead
and the dot, with $\Gamma\rightarrow0$ and
$\Nphi\rightarrow\infty$ while maintaining
$\Gamma\Nphi=2\pi\hbar/\Delta\tau_\phi$. \cite{brouwer97}  The
S-matrix is then not drawn from the CSE, and simple analytical
results in the spin-orbit coupled system are challenging.  We
study this model numerically and find that for fixed
$\tau_\phi$, it gives qualitatively similar results to the
simpler model described above; in particular it also gives a
nonzero spin current when $\NR=1$ (not shown).
Without a microscopic model of dephasing, it is possible that
this dephasing induced spin current with $\NR=1$ is an artifact
of third lead dephasing models, but all three variants of third
lead dephasing discussed here show this effect, so dephasing
gives a loophole for producing spin currents even when $\NR=1$.

Returning to the single dephasing lead with $\Gamma=1$, we
estimate $\av{\pol^2}$ as above, where we modify $\av{\prob}$
to include the dephasing lead.  Using Eqs.\ \ref{eq:2S} and
\ref{eq:win_phi_avg}, this gives
\begin{alignat}{10}\label{eq:p2_dephasing}
    \av{\prob}\approx\frac{2\NR\Ntotp}{\Ntot(2\Ntotp-1)},
\end{alignat}
We estimate $\av{\pol^2}\approx\av{(\cur)^2}/\av{\prob}^2$,
using Eqs.\ \ref{eq:n_phi} and \ref{eq:p2_dephasing}.
Comparison of these approximations to numerical evaluations is
shown in Fig.\ \ref{fig:s2_phi}. Again we find that the
numerical and analytical results agree qualitatively, except
when $\Nphi=\NR=1$.

\section{Finite Temperature}

If the temperature $T>\Delta$, the polarization will be further
suppressed by electrons of different energy feeling
uncorrelated scattering matrices. This effectively increases
the number of orbital channels, which decreases the residual
polarization. We consider unpolarized incoherent flux from the
left lead at temperature $T$. Adapting Datta, \cite{datta} we
take $\win(\epsilon)=-\frac{\partial \fermi}{\partial \epsilon}
\frac{1}{2\NL}
\left(\begin{smallmatrix}\openone_\NL&&\\&0_\NR&\end{smallmatrix}\right)$,
where $\fermi(\epsilon)$ is the Fermi distribution. If the
scattering matrix for particles of energy $\epsilon$ is
$S(\epsilon)$, then
$\wout(\epsilon)=S(\epsilon)\win(\epsilon)S^\dagger(\epsilon)$.
We approximate $S(\epsilon)$ as correlated only within energy
intervals of scale $\Delta$ (see Ref.~\onlinecite{huibers98}
for an equivalent treatment). That is, we take
\begin{align}
  \av{S_{ab}(\epsilon) S^\dagger_{cd}(\epp)}
  &= \Delta  \delta(\epsilon-\epp)
  \av{S_{ab}(\epsilon)S_{cd}^\dagger(\epsilon)},
\end{align}
and
\begin{align}\label{eq:S_correlator}
    \Big<S_{ab}(\epsilon)&S_{cd}(\epp)   S^\dagger_{ef}(\epsilon)S^\dagger_{gh}(\epp)\Big>\nonumber\\
    &=\av{S_{ab}(\epsilon)S_{ef}^\dagger(\epsilon)}\av{S_{cd}(\epp)S^\dagger_{gh}(\epp)}\\
    &+\Delta\delta(\epsilon-\epp)
    \av{S_{ab}(\epsilon) S_{cd}(\epsilon)S^\dagger_{ef}(\epsilon)S^\dagger_{gh}(\epsilon)},\nonumber
\end{align}
which are valid only for $T\gg\Delta$, which is often true for
chaotic quantum dots. For $T\approx \Delta$,
$\av{S(\epsilon)S^\dagger(\epsilon)}$ can be calculated using
the random Hamiltonian method. \cite{Verbaarschot85}

We need an average over a new function,
\begin{align*}
    h(\epsilon,\epp)=\fermi^\prime(\epsilon)\fermi^\prime(\epp)
    \Tr[A S(\epsilon)B S^\dagger(\epsilon)]
    \Tr[C S(\epp) D S^\dagger(\epp)],
\end{align*}
where $A B=A D=C B=C D=0$ and
$\fermi^\prime=\partial\fermi/\partial\epsilon$. We evaluate
the average of $h$ with the $\Ntot\times\Ntot$ quaternion
matrix $S(\epsilon)$ chosen from the CSE along with
Eq.~\ref{eq:S_correlator}, giving,
\begin{widetext}
\begin{alignat}{10}
  \int\rd{\epsilon}\rd{\epp}\av{h(\epsilon,\epp)}
  &=\frac{4}{(2\Ntot-1)^2}\Tr(A)\Tr(B)\Tr(C)\Tr(D)\nonumber\\
  +\frac{\Delta}{\Lambda}\int\rd{\epsilon}
    \fermi^\prime(\epsilon)^2
    \big\lbrace&(\Ntot-1)[4\Tr A\Tr B\Tr C \Tr D
    +\Tr(A C)\Tr(B D)]
    - \Tr A\Tr C \Tr(BD)-\Tr(AC)\Tr B \Tr D\big\rbrace \nonumber\\
  &= \frac{4}{(2\Ntot-1)^2}\Tr(A)\Tr(B)\Tr(C)\Tr(D)\nonumber\\
  +\frac{\Delta}{6T\Lambda}
    \big\lbrace(\Ntot-1)&[4\Tr A\Tr B\Tr C \Tr D
    +\Tr(A C)\Tr(B D)] - \Tr A\Tr C \Tr(BD)-\Tr(AC)\Tr B \Tr
    D\big\rbrace.
    \label{eq:4S_E}
\end{alignat}
\end{widetext}

Using Eq. \ref{eq:4S_E} in place of Eq.\ \ref{eq:4S}, we
evaluate $\av{(\cur)^2}$ as above, which simply multiplies Eq.\
\ref{eq:n_msq_unpolarized} by $\frac{\Delta}{6T}$. Also,
$\av\prob$ is unaffected by temperature, so Eq.~\ref{eq:s2} is
also multiplied by $\Delta/6T$.

When dephasing and temperature are both included, the
scattering matrix is correlated on the scale of the level
broadening, $\Delta(1+\Nphi/2)$, \cite{huibers98} so we replace
$\Delta$ in Eq.\ \ref{eq:S_correlator} by $\Delta(1+\Nphi/2)$.
 Eq.\ \ref{eq:n_phi} is then multiplied by
$\Delta(1+\Nphi/2)/6T$, and Eq.\ \ref{eq:p2_dephasing} is
unchanged.

\section{Discussion}
This spin polarization should be able to be produced and
detected experimentally. Even quantum dots in n-type
GaAs/AlGaAs heterostructures have been observed to have
sufficiently strong spin-orbit coupling to approach the RMT
symplectic limit. \cite{zumbuhl02,zumbuhl05} If the spin-orbit
coupling is not strong enough for the S-matrices of the dot to
be drawn from the CSE, the spin polarization predicted here
will be reduced but should still be present. In a given
material with fixed spin-orbit coupling strength, a
sufficiently large quantum dot will be well described by the
CSE, with a possible increase in dephasing rate as the dot size
increases.

At zero temperature, our discussion has assumed that all
electrons passing through the dot see the same S-matrix, which
is valid when the applied potential difference is less than the
mean level spacing $\Delta$. Since $\Delta\propto A^{-1}$, as
the dot area is increased to approach the strong spin-orbit
limit, the window of voltages where these results apply
shrinks. The effects predicted in this paper are most likely to
be observable in a material with inherently strong spin-orbit
coupling, such as p-type III/V heterostructures. Note that
$\av{(\cur_\mu)^2}=\av{(\cur)^2}/3$ and
$\av{\pol_\mu^2}=\av{\pol^2}/3$, so if a measurement technique
or application is only sensitive to spin polarization along a
particular axis, then the rms predictions for the
$\mu$-component of the polarization and spin conductance are
only $\sqrt3$ times smaller than the results stated above.

We have shown that quantum dots with spin-orbit coupling can
generate spin polarized currents without magnetic fields or
ferromagnets, except with only one outgoing channel and TRS,
when such a device cannot produce a spin current. Mesoscopic
fluctuations can be large enough to give appreciable spin
currents in devices with a small number of propagating
channels.  Even if the spin-orbit coupling is weak, a
sufficiently large device will show these effects.

\begin{acknowledgments}
  We acknowledge helpful discussions with Caio Lewenkopf,
  Emmanuel Rashba, and particularly Ari Turner, who suggested
  the method of choosing matrices from the CSE. We note the use
  of the Quaternion Toolbox for MATLAB, created by S.\ J.\
  Sangwine and N.\ Le Bihan. This work was supported in part by
  the Fannie
 and John Hertz Foundation and NSF grants PHY-0646094 and DMR-0541988.
\end{acknowledgments}

\bibliography{SFET,CubicDressRefs}

\end{document}